\documentclass[preprint2,twoside]{hwo}

\usepackage{graphicx}
\usepackage{upgreek}
\usepackage{chemformula}
\usepackage[version=4]{mhchem} 

\input{hwo.h}

\begin{document}

\title{\textbf{\LARGE Characterizing the Dynamics and Chemistry of Transiting Exoplanets with the Habitable World Observatory}}
\author {\textbf{\large Hannah R. Wakeford$^{1}$ and Laura C. Mayorga$^{2}$}}
\affil{$^1$\small\it University of Bristol, School of Physics, HH Wills Physics Laboratory, Tyndall Avenue, Bristol, BS8 1TL, UK}
\affil{$^2$\small\it Johns Hopkins Applied Physics Laboratory, Laurel, MD, 20723, USA}

\author{Contributing Authors: Joanna K. Barstow$^{3}$, Natasha E. Batalha$^{4}$, Ludmila Carone$^{5}$, Sarah L. Casewell$^{6}$, Theodora Karalidi$^{7}$, Tiffany Kataria$^{8}$, Erin M. May$^{2}$, Michiel Min$^{9}$} 
\affil{$^{3}$ School of Physical Sciences, The Open University, Milton Keynes, UK} 
\affil{$^{4}$ NASA Ames Research Center, MS 245-3, Moffett Field, CA 94035, USA} 
\affil{$^{5}$ Space Research Institute, Austrian Academy of Sciences, Schmiedlstrasse 6, A-8042 Graz, Austria}
\affil{$^{6}$ School of Physics and Astronomy, University of Leicester, Leicester LE1 7RH, UK}
\affil{$^{7}$ Department of Physics, University of Central Florida, 4111 Libra Drive, Orlando, FL 32816, USA} 
\affil{$^{8}$ NASA Jet Propulsion Laboratory, California Institute of Technology, Pasadena, CA, USA}
\affil{$^{9}$ SRON Netherlands Institute for Space Research, Niels Bohrweg 4, 2333 CA Leiden, the Netherlands}

\author{\footnotesize{\bf Endorsed by:}
Munazza Alam (STScI), Natalie Allen (Johns Hopkins University), Reza Ashtari (JHU-APL), Katherine Bennett (Johns Hopkins University), Jayne Birkby (University of Oxford), V. Abby Boehm (Cornell University), Kara Brugman (Arizona State University), Aarynn Carter (STScI), Katy Chubb (University of Bristol), Jamie Dietrich (Arizona State University), Luca Fossati (Space Research Institute, Austrian Academy of Sciences), Siddharth Gandhi (University of Warwick), Caleb Harada (University of California, Berkeley), Cenk Kayhan (Kayseri University), Eliza Kempton (University of Chicago), James Kirk (Imperial College London), Thaddeus Komacek (University of Oxford), Adam Langeveld (Johns Hopkins University), Eunjeong Lee (EisKosmos (CROASAEN), Inc.), Nataliea Lowson (University of Delaware), Benjamin Montet (University of New South Wales), Iain Neill Reid (STScI), Gaetano Scandariato (INAF), Melinda Soares-Furtado (University of Wisconsin–Madison), Daniel Valentine (University of Bristol), Vincent Van Eylen (University College London), Peter Wheatley (University of Warwick).
}

\begin{abstract}
The primary scientific objective of this Habitable Worlds Observatory (HWO) Science Case Development Document (SCDD) is to measure planetary rotation rates of transiting exoplanets to determine the structure, composition, circulation, and aerosol properties of their planetary atmospheres. For this analysis, \textit{HWO} would obtain spectroscopic phase curves for planets with orbital periods of 5--20+ days, to assess tidal locking radius assumptions. Extending phase curve studies out to longer orbital periods than accessible with current and near-future telescopes will enable detailed investigation of atmospheric structure, composition, and circulation for planets that are much cooler than the more highly irradiated planets accessible with \textit{JWST} phase curve observations (i.e., T$_\mathrm{eq}$\,$<$\,500\,K for HWO versus 1400\,K\,$\le$\,T$_\mathrm{eq}\,\le$\,2600\,K for \textit{JWST}). Broad wavelength coverage extending from the UV to the NIR would capture both reflected light and thermal emission, enabling \textit{HWO} to conduct comprehensive characterization of planetary atmospheres. UV observations would probe high altitudes, thereby providing valuable insights into atmospheric (dis)equilibrium, aerosol properties, and the effects of photochemical processes on atmospheric composition. We also discuss the role of polarimetry in the classification of aerosols and the associated role they play in the atmospheric energy budget that directly ties them to the chemistry and circulation structure of the atmosphere. 
\\
\\
\end{abstract}

\vspace{2cm}

\section{Science Goal}
The key science question for this Habitable World Observatory (\textit{HWO}) science case is:
\textit{\textbf{At what point does super-rotation become inefficient in tidally locked exoplanets and how does this impact the multi-dimensional thermodynamics and chemistry of the atmosphere?}}

The Astro 2020 decadal detailed a range of core scientific questions and goals for the scientific community over the next decade. One of those, ``\textit{What are the properties of individual planets, and which processes lead to planetary diversity?}'' will likely continue into the following decades and requires a long term commitment through multiple observatories to truly answer. This goal has set the foundation for a number of missions such as \textit{JWST}’s Directors Discretionary Exoplanet initiative \citep{JWSTDDT2024}, as well as the ESA \textit{Ariel} mission dedicated to surveying the atmospheres of 100's of transiting exoplanets \citep{Tinetti2022EPSC}. Transiting exoplanets currently represent the largest pool of exoplanets amenable to detailed characterization, enabling true statistically robust population studies. 

One foundational question when looking at close-in transiting exoplanets is: \textit{What controls the balance between advective and radiative timescales in tidally locked giant exoplanets across the population?} Advective timescales define the time taken for heat to be actively transported around the planet from the sub-stellar point, which is the part of the atmosphere closest to the star; advective timescales depend strongly on the structure of the global and local climate which shape the momentum transport throughout the atmosphere. Radiative timescales are defined by the time it takes for the incoming radiation to be reflected or emitted by the planet out into space; radiative timescales are strongly dependent on the chemistry of the atmosphere and the compositions, structure and location of aerosols (clouds and hazes). The balance between advective and radiative processes tells us about the composition, thermal structure, and circulation structure of the planet's atmosphere, in turn informing us about interactions between the fundamental physics and chemistry \citep[see][]{Showman2010book}. 
In part, the relationship between these two factors has and is being answered through work with ground- and space-based instrumentation measuring the thermal dayside emission, chemical composition from transmission studies, and longitudinal distribution from close-in tidally locked exoplanet phase curves \citep[e.g.,][]{Kreidberg2014ApJ,Sing2016Nature,Ehrenreich2020Nature,Mansfield2021NatAs,CarterMay2024NatAs,Challener2024ApJ,Valentine2024AJ}. However, current studies lack a number of aspects: limited or no access to short wavelengths where reflected and scattered light become dominant ($<$1\,$\upmu$m); poor precision and spectral resolution in the UV–optical, preventing the detection of specific atoms and molecular signatures responsible for photochemical reactions; and short observing times that limit the ability to measure extended phase curves for longer period transiting exoplanets. Fundamentally, enabling advancement in any one of these can improve the information we have on close-in transiting exoplanets, but only with a drive towards addressing all of these issues can we make a paradigm shift in our understanding of the connection between thermal and chemical drivers in an irradiated atmosphere as a population and get a true look at changing climates and structures relating to their environments. Critically, current observations are limited in the observatory baseline, not necessarily in what is possible (\textit{Ariel} for example should have the ability to observe out to 5 day orbits), but in what is funded and supported (for example selection by the Time Allocation Committee for \textit{JWST} studies). Longer observing baselines allow access to a wider population of transiting exoplanets and one of the future goals of the field is to measure: \textit{How rotation, based on empirical evidence across the tidal locking boundary, impacts the dynamics and therefore radiative and chemical profile of an atmosphere?}. 

\begin{figure}[h]
    \centering
    \includegraphics[width=1\linewidth]{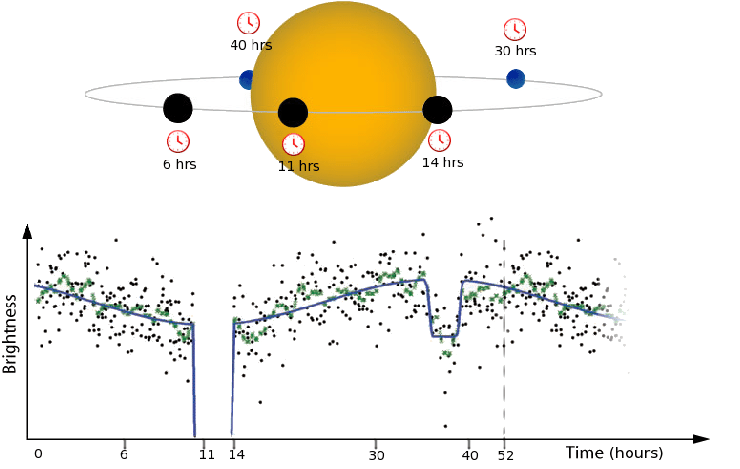}
    \caption{Optical phase curve of HAT-P-7b from the Kepler Space Telescope \citep{Borucki2009Sci} showing the flux variations as a function of time as the planet orbits the star. This phase curve consists of 52+ hours of observing time and revealed a strong day-to-night contrast for the planet suggestive of poor advection of flux. \textit{Image compilation from \citet{Tinetti2012EcHO} EChO mission concept proposal}.}
    \label{fig:phase-curve}
\end{figure}

\section{Science Objective}
Our primary science objective is to measure the climate of a large number of transiting exoplanets at different orbital periods, eccentricities, stellar irradiation types, and radius/mass to assess the impact of system environment and planetary properties on their circulation and chemistry.

\textit{HWO} will be critical in finishing the scientific journey that previous flagship observatories such as \textit{Hubble}, \textit{Spitzer} and \textit{JWST} have started. To spatially map the changing climate of a transiting exoplanet we require observations through complete phase curve observations at high time cadence, precision, and over the reflected light to thermal emission wavelength range (reflected $\sim$0.2 -- 1\,$\upmu$m; thermal $>$0.5\,$\upmu$m - stretching to longer wavelengths for cooler planets). To fully address population level questions for giant exoplanets will require a statistically significant sample of phase curves to be measured for planets with different properties such as eccentricity, orbital period, stellar spectral type, metallicity/composition and mass/radius.

\subsection{Orbital Phase Curves}
For transiting exoplanet studies, orbital phase curves represent the most comprehensive demands on the telescope architecture, requiring long stable pointing on the order of days and data bandwidth to obtain continuous repeated measurements over seconds or minutes for the whole observing time. As such, if the scientific and technical specifications necessary to observe long time series phase curves can be met by \textit{HWO}, then any observations of transiting exoplanets should be in the scope of the telescope capabilities as they would represent a subset of a phase curve. 

\begin{figure*}[ht]
    \centering
    \includegraphics[width=1\textwidth]{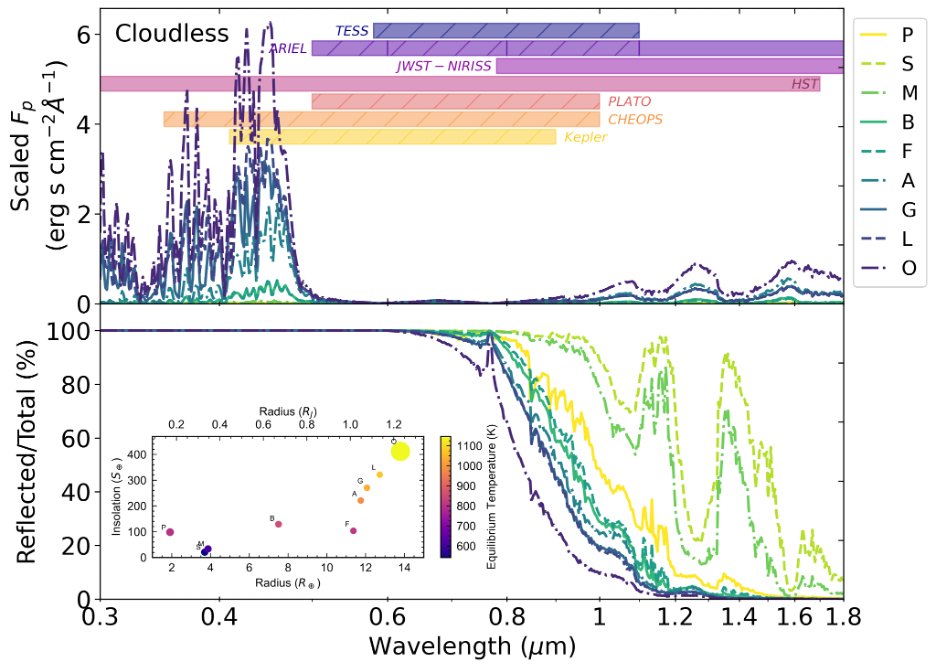}
    \caption{To understand where the measured flux from the planet is coming from we need to capture both reflected and thermal components of the light from the planet. This requires broad wavelength coverage from $<$0.5 to $>$1 micron. Predicted planetary reflected signal vs. thermal emission for archetypal planets (letters correspond to planet properties shown in the inset plot) identified by \citet{Mayorga2019AJ}. Top: the total planetary flux as measured in flux units. Bottom: the percent contribution of reflected light to the total flux as a function of wavelength (R~300). Colored bands in the top panel show the photometric (hashed) and spectroscopic (solid) coverage from current and future facilities over this wavelength range. \textit{Figure courtesy of L. C. Mayorga.}}
    \label{fig:mayorga2019}
\end{figure*}

Figure\,\ref{fig:phase-curve} shows an example phase curve observation from the Kepler Space Telescope consisting of over 50 hours of observation \citep{Borucki2009Sci}. The phase curve can be used to horizontally map the longitudinal structure of the planet's flux through measurements of its amplitude and peak flux output (known as the hotspot offset). The hotspot offset is driven by temperature differentials between the permanent day side to the permanent night side, which can be measured from the amplitude of the phase curve from transit to eclipse, driving strong super-rotating (eastward) equatorial winds. As a result, the hottest hemisphere is offset relative to the day side hemisphere. In phase curve observations we observe this as the `peak' planet emission occurring at a different time from when the day side hemisphere is directly facing the observer. The time scale of this hotspot offset gives us insights into the strength of the winds, which when combined with the measured temperature difference from day to night provides information on the effectiveness of the planet’s circulation and to some extent its overall structure.  

Hot Jupiters, close-in giant exoplanets which are highly irradiated by their host star, are dominated by external heating, as opposed to internal heating as seen in brown dwarfs or Jupiter. The overall atmospheric structure of hot Jupiters is driven by this extreme external forcing, and is characterized by large-scale horizontal structures such as the super-rotating jet that transports heat from the day side to the night side, and a vertically stratified atmosphere with a temperature inversion where strong UV-optical absorbers are present. As the orbit of giant exoplanets from their stars increases their temperature will decrease (if considered around the same type of star) this combination of changing irradiation environment and longer rotation period will impact both the vertical and horizontal transport in the atmosphere. As transiting giant exoplanets transition from hot, to warm or cool Jupiters, going from thousands ($T_\mathrm{eq}\,>$\,1500\,K) to hundreds ($T_\mathrm{eq}\,<$\,500\,K) of Kelvin in temperature, they may develop Jupiter-like structures where rapid planetary rotation coupled with a reduction in external forcing results in multiple alternating bands of east/west circulation, or turbulence due to large internal heat may dominate the vertical structure as seen in some brown dwarfs \citep[see e.g.,][]{Showman2020SSRv,Showman2021book}.

\subsection{The need for Spectroscopic Measurements}
While \textit{Kepler} used a photometric bandpass in the optical-IR to measure broadband inferences about the atmosphere of a handful of optimal cases, to make accurate and informed inferences about the composition and structure of the planetary atmosphere, and therefore the nature of physics and chemistry, requires spectroscopic observations. 
The chemistry, often driven by the temperature, plays a significant role in the emergence of flux from a planetary atmosphere. The UV in particular is a driver for atmospheric chemistry through photochemical reactions and the high energies necessary to drive equilibrium or disequilibrium processes. Spectroscopy is vital to measuring the abundances of different material in an atmosphere, each atom and molecule imprinting a unique signature on the spectrum through absorption, emission, or scattering effects. Figure\,\ref{fig:mayorga2019} shows the balance of reflected to thermal light from a planetary atmosphere, the spectra displaying different amplitudes of absorption and emission depending on the planet-star system parameters \citep[see][]{Mayorga2019AJ}. The photometric coverage of missions such as \textit{Kepler} and \textit{CHEOPS} span a wide photometric wavelength stopping at the critical point where reflected and thermal light switch dominance in the output flux and therefore no information can be gained on the impacting chemistry, due to the lack of spectroscopic instrumentation, on the overall energy of the atmosphere. While the ESA Ariel Mission has the potential to build a foundation of knowledge on optical photometric phase curves through three photometric bands that cover 0.5\,--\,0.6\,$\upmu$m, 0.6\,--\,0.8\,$\upmu$m, and 0.8\,--\,1.1\,$\upmu$m these wavelengths are still unable to reliably sample the reflected light component of a large number of target atmospheres and lack the needed spectral resolution to resolve critical UV species. \textit{Ariel} may prove key in defining the target list for spectroscopic follow-up with \textit{HWO} with a present list of 136 known giant planet targets and a nominal ability to observe a single target up to 5-days continuously \citep{Tinetti2022EPSC}. To categorically disentangle the reflected from thermal component of the atmosphere both must be measured, this requires spectroscopic coverage from the UV capturing the reflected component and key atomic species such as Fe, Mg, Na, and K, to the near-IR ($>$1.6\,$\upmu$m) to capture absorption from molecular species like \ce{H2O}, \ce{HCN}, \ce{CH4}, \ce{NH3} \citep{Moses2011ApJ} which interact thermally with the atmosphere and are key metallicity indicators. Additionally, pushing out to 5\,$\upmu$m then gives information on \ce{CO2} and \ce{CO} in the atmosphere as well as the secondary bands of sulfur-bearing photochemically generated species to better determine the C/O, S/O, Si/O and other chemical ratios that tie back to formation conditions \citep[e.g.,][]{Oberg2011ApJ,Mordasini2016ApJ}. The more complete the wavelength coverage in both directions about the optical the more complete the information. 

\begin{figure}[ht]
    \centering
    \includegraphics[width=1\linewidth]{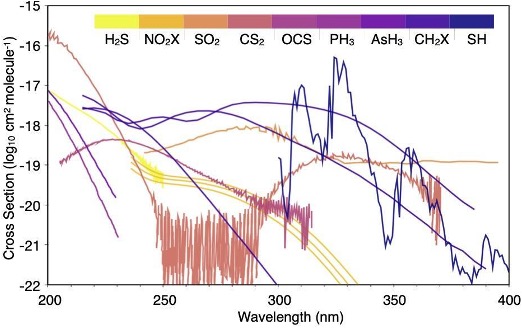}
    \caption{Photoabsorption cross sections for a range of molecules in the UV–optical wavelength range that represent key photolysis species responsible for photochemical chains and production of photochemically generated species. Here \ce{NO2X} and \ce{CH2X} represent the baseline cross section for any species based on this chemical structure. \textit{Figure courtesy of Natasha E. Batalha and reproduced from \citet{Wakeford2022hst} (HST GO-17183, HUSTLE treasury program).}}
    \label{fig:HUSTLE-flux}
\end{figure}

\begin{figure*}[ht]
    \centering
    \includegraphics[width=1\textwidth]{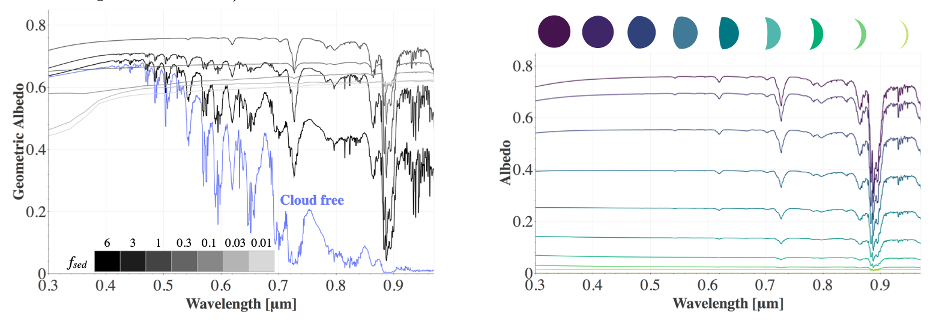}
    \caption{The impact of clouds on the geometric albedo of a giant planet atmosphere. Left: Representative albedo spectra showing the effect of varying cloud profiles for a 1$\times$ Solar composition planet located 5 AU from a Sun-like star (gravity\,=\,25\,m/s$^2$). (1) large fsed's (vertical settling component of the atmosphere) create vertically thin, optically thin clouds and vice versa, (2) Clouds increase atmospheric brightness toward 1\,$\upmu$m. Right: Representative albedo spectra showing the effect of phase when clouds are also present for an fsed\,=\,3. The higher phase (from full phase\,=\,0) observations decrease the overall brightness of the directly imaged planet. \textit{Figure from \citet{Batalha2018AJ}.} }
    \label{fig:clouds}
\end{figure*}

\subsubsection{Dis-equilibrium Processes}
The UV is critical not just for assessing the reflected light component of an atmosphere but to also assess disequilibrium processes in planetary atmosphere because, 1) the millibar to microbar pressure levels accessed by UV wavelengths are where processes such as photochemistry, ion chemistry, and other photochemical processes heavily shape the overall composition of the atmosphere, and 2) the chemical species which drive these processes generally have large cross sections in the UV (thus are drivers of processes in the UV). In Figure\,\ref{fig:HUSTLE-flux} we show a series of potential UV absorbers between 200\,--\,400\,nm that could be detected by transmission or reflected light observations - OH, CH, CN \citep{Lavvas2021MNRAS}, \ce{CS2} \citep{Grosch2015JQSRT}, \ce{PH3}, OCS and SH \citep{Orphal2003JQSRT}. Many of these UV candidate absorbers impact other astrophysical environments, such as \ce{PH3} which is important for Saturn's bulk chemistry \citep[e.g.,][]{Chen1991JGR}, and \ce{CS2} which was discovered in Neptune's atmosphere via ALMA data suggestive of deposition by a large cometary impact \citep[e.g.,][]{Moreno2017MNRAS}. As seen in the solar system, it is likely that key chemical species that play critical roles in shaping global scale chemical and radiative processes remain undetected at optical and IR wavelengths, but strongly shape the M-NUV portion of planetary atmospheric spectra. Fundamentally, UV wavelengths probe atmospheric physical and chemical processes that could easily be overlooked or misinterpreted when considering the optical and IR alone and it will be critical that any future instrument designed to capture a complete picture of planetary atmospheres has UV spectroscopic capabilities (see also Figure\,\ref{fig:mayorga2019}). 

\subsection{Aerosols}
One of the main atmospheric components that drive local and global climate are aerosols. Aerosols, solid or liquid particles suspended in the gas, can effectively reflect or absorb radiation at different wavelengths changing the energy balance of the atmosphere. In general these are defined in two categories, clouds formed through condensation processes that are reversible, and hazes through photochemical processes that are irreversible\footnote{However, we note that in exoplanet literature haze is often used to describe small particle scattering aerosols and clouds for large particle uniform opacity aerosols regardless of composition or formation mechanism}. The presence and formation of aerosols is dictated by the conditions of that atmosphere and in turn they then have an impact on those conditions creating a delicate balance in feedback and forcing environments. All atmospheres have aerosols to some extent. In transiting exoplanet atmospheres these have been seen to mute molecular absorption featured in the IR and cause enhanced scattering in the UV-optical \citep[e.g.,][]{Pont2008MNRAS,Sing2016Nature,Fu2017ApJ,Helling2019AREPS,Barstow2021AG,Gao2021JGRE}. The location (horizontal and/or vertical), composition, and structure of aerosols will have varying impacts on the emergent flux of the planet changing the day-to-night temperature contrast and potentially shifting the hotspot. Only through spectroscopic measurements, however, can these effects be disentangled from global gaseous atmospheric properties. Figure\,\ref{fig:clouds} shows the impact of clouds on the measured geometric albedo of an atmosphere based on settling properties (fsed) from 0.01 to 6 moving from optically thick clouds to optically thin clouds \citep{Batalha2018AJ}. This shows that clouds can increase the brightness of a planet in reflected light compared to a cloud free scenario, and that for a cloudy planet the phase curve would be highly modulated by the albedo of the clouds.   
\begin{figure*}[ht]
    \centering
    \includegraphics[width=1\textwidth]{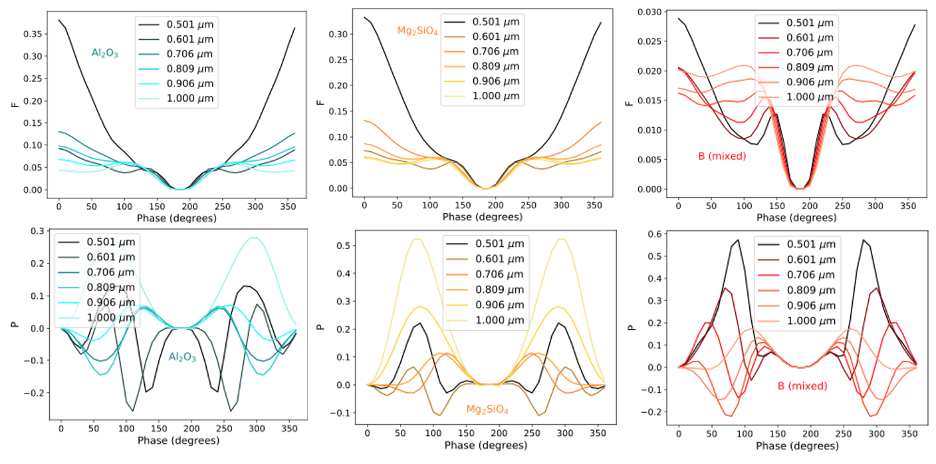}
    \caption{The vector nature of polarization makes it sensitive to microphysical properties of the clouds. The phase curve of an exoplanet dominated by \ce{Al2O3} (left panels), \ce{Mg2SiO4} (middle panels) or mixed clouds (right panels) shows a stronger variability in polarized light (bottom panels) than unpolarized light (top panels). Also note the large variability in the zero points (phase angles where P\,=\,0) that allows us to better distinguish the different scenarios. \textit{Figure adapted from  \citet{Chubb2024MNRAS}.}}
    \label{fig:chubb}
\end{figure*}

\subsubsection{Polarization}
Another key aspect to cloud studies in an exoplanet atmosphere when considering short wavelength measurements is polarization. Polarimetric observations complement unpolarized light phase curve and spectroscopy observations allowing for a more accurate characterization of an exoplanet atmosphere. Due to the vector nature of polarization, polarimetric phase curves at different wavelengths vary strongly with variations in the properties of the atmosphere. Polarimetric phase curves thus, can help constrain atmospheric properties like the refractive index of clouds, cloud altitude, cloud optical thickness, spatial distribution of clouds in an atmosphere etc. breaking a number of degeneracies that unpolarized light observations have \citep[e.g.,][]{Stam2004AA,Bailey2012MNRAS,Karalidi2013AA,Stolker2017ApJ,Bailey2018MNRAS,Chubb2024MNRAS}. Figure\,\ref{fig:chubb} shows simulated phase curves of an exoplanet with different dominant cloud species and the measured effect at different wavelengths, once again demonstrating the need for spectroscopic observations to distinguish between atmospheric compositions and in this case the role of spectroscopic polarimetry in assessing composition and structure of the clouds. 

Ground-based telescope efforts to detect visible wavelength polarization from hot Jupiters WASP-18b \citep{Bott2018AJ} and HD\,189733b \citep{Wiktorowicz2015ApJ,Bott2016MNRAS} set upper limits to the reflected light polarization of these worlds at a level of 40 to 70 ppm. Specifically, the upper bound to polarization from the hot Jupiter exoplanet WASP-18b provided by a non-detection ruled out a high-altitude Rayleigh scattering haze. Models of reflected light polarimetry of tidally locked giant exoplanets suggest an expected polarimetric signal on the order of 0.1 to 40 ppm \citep{Bailey2018MNRAS,Chubb2024MNRAS}, strongly dependent on the nature of clouds, optical thickness and the patchiness in the atmosphere (see Figure\,\ref{fig:chubb}). Some species of cloud expected at these hot temperatures are poor reflectors of light, suggesting that some more reflective species (e.g., \ce{KCl}, \ce{H2O}) expected in cooler atmospheres at greater separations from their star may produce stronger signals.  Reaching such levels is near the abilities of modern day polarimeters. Polarimetric phase curves at levels of a few ppm will be crucial for breaking degeneracies in the characterization of clouds in the atmospheres of exoplanets and accurately mapping the distribution of clouds across the atmosphere, which will allow a better characterization of atmospheric circulation and chemistry. \\

\begin{figure}[h]
    \centering
    \includegraphics[width=1\linewidth]{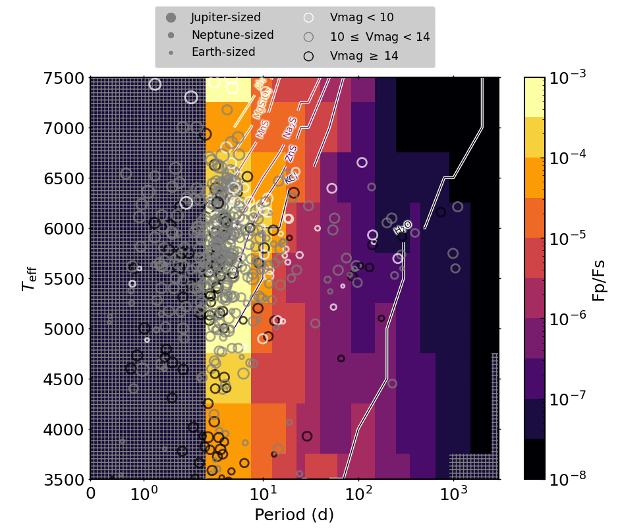}
    \caption{Exoplanet transit population as a function of stellar effective temperature and orbital period on a backdrop of predicted planet-to-star contrast ratios (averaged from 0.3-1.0 micron) and cloud contours. Circle size indicates planetary radius and circle color indicates stellar V magnitude. Contours indicate which condensate (if any) should be forming at the highest altitude in the theoretical atmosphere. \textit{Figure courtesy of L. C. Mayorga.}}
    \label{fig:mayorga_dist}
\end{figure}
\noindent{}The population of exoplanets amenable to high precision phase curve information is potentially in the 100s–1000s. Figure\,\ref{fig:mayorga_dist} shows the planet to star contrast ratio computed for the transiting exoplanet population using a 1D radiative-convective equilibrium model assuming a solar metallicity, M$_\mathrm{p}$\,=\,1\,M$_\mathrm{J}$, R$_\mathrm{p}$\,=\,1\,R$_\mathrm{J}$, that the atmosphere is in chemical equilibrium, the planet is in a circular orbit for their given orbital period around a host star with a given stellar effective temperature. We note that this model used is invalid for ultra-hot Jupiters and thus this region is shaded out in the diagram. The figure demonstrates that across this range of parameter space, the atmospheres of these Jupiter-sized planets could be quite different and the giant planet population is even more varied in mass and radius.

Critically, based on our outlined science goal, and the necessary observational set-up required to achieve it, this case outlines a strategy that will allow \textit{HWO} to perform any transiting exoplanet or time series observation (short single eclipse or transit observations, time series stellar-host star observations, etc.) to high precision and accuracy.

\section{Physical Parameters}
For these measurements we will need to consider transiting exoplanets with the following known physical parameters:
\begin{itemize}
    \item Orbital period,
    \item Host star temperature,
    \item Planetary equilibrium temperature,
    \item Mass/gravity/radius, and
    \item Eccentricity.
\end{itemize}

The range of orbital periods is the first order physical parameter that defines the length of observation needed to address our science goal: empirically mapping the tidal locking radius to understand the drivers of climate. The combination of orbital period with stellar temperatures will define the temperature range of planets that can be assessed. 

These would then form the basis of a population to determine the following:
\begin{itemize}
    \item Atmospheric circulation and energy balance through hotspot offset and day-to-night temperature contrast.
    \item Atmospheric composition and structure from spectroscopic measurements.
    \item Cloudiness of the atmosphere and the aerosol properties from scattering and polarimetry.
\end{itemize}
With the aim to assess the 
\begin{itemize}
    \item Planetary rotation period, atmospheric chemistry, and impact on the atmospheric structure.
\end{itemize}

\begin{table*}[ht!]
    \centering
    \caption[Performance Goals]{Phase curve number, range of planetary period, temperature, and aerosol property benchmarks for \textit{HWO}.}
    \label{tab:performance}
    \begin{tabular}{|p{0.15\linewidth}|p{0.2\linewidth}|p{0.15\linewidth}|p{0.15\linewidth}|p{0.2\linewidth}|}
        \noalign{\smallskip}
        \hline
        \noalign{\smallskip}
        {Physical Parameter} & {State of the Art} & {Incremental Progress} & {Substantial Progress} & {Major Progress} \\
        \noalign{\smallskip}
        \hline
        \noalign{\smallskip}
        Number of Objects &
        Roughly 20, over lifetime of \textit{JWST}, potential survey of 10's with \textit{Ariel} &
        100+ available but would be able to do 9 with 5\% telescope time for 5\,year mission &
        $\sim$300 available, can do 5 with 5\% time &
        $\sim$1000+ available and target selection will need to be carefully assessed to address trends. \\
        \hline
        Orbital Period & 
        $<$\,2 days \textit{JWST}, $<$\,5 days Ariel &
        Up to 10 days &
        Up to 20 days &
        $>$\,20 days this will represent a clear major breakthrough for \textit{HWO} \\
        \hline
        Planet Temperature Range/Distribution &
        Defined by the combination of period and stellar Teff. \textit{JWST} range for phasecurves $\sim$1400--2600\,K &
        Gas giant planet in phase curve down to $<$\,1000\,K &
         Gas giant planet in phase curve $<$\,700\,K &
        $<$500\,K \\
        \hline
        Composition and structure of aerosols &
        Measurement of IR absorption with \textit{JWST} in emission and transmission, HST scattering slope in transmission &
        Distribution of clouds spatially relating to circulation structure &
        Spatial location and size distribution of cloud species &
        Determination of aerosols without IR absorption via UV spectroscopy. Determination of fractal shape of aerosols from polarimetry. \\
        \noalign{\smallskip}
        \hline
    \end{tabular}
\end{table*}

Continuous phase curve observations for longer period ($>$\,5 day) transiting exoplanets will require a significant amount of telescope time. To address the feasibility of this as a population study we calculate an example time case considering a 5-year nominal mission and a minimum of 5 planetary phase curves measured. 

Example time cases: To obtain 5 phase curves on 5 or 10 day orbits with $\sim$250 hours of observation each this would total 1,250 hours which is 14\% of a year. Assuming the telescope's nominal mission is 5 years this would equate to $\sim$3\% of total telescope time. To obtain 5 phase curves of planets on 20 day orbits this would be 5.5\% of total mission lifetime. These values assume 100\% efficiency of observations, if we instead assume 75\% telescope efficiency (approximately what is achieved for phase curves with \textit{JWST}) this would be 7.4\% for 5 planets on 20 day orbits and 3.8\% for 5 planets on 10 day orbits. A carefully designed survey to sample the untapped phase curves of transiting exoplanets will be necessary to ensure the science goal is achieved (see \citealt{Batalha2023AJ} for examples of setting up statistically valid surveys).

As current state-of-the-art facilities such as \textit{JWST} and \textit{Ariel} are limited in both wavelength coverage and the length of orbital period accessible in phase curve observations, the priority for a phase curve survey with \textit{HWO} will be to push towards phase curves for planets with longer periods of $>$\,5 days which cannot be performed with \textit{JWST} or \textit{Ariel}. We will also recommend that we obtain phase curve data for a few existing \textit{JWST} and \textit{Ariel} targets, to allow comparison with existing observatories and to establish the link between UV, optical and infrared phase variations. Once this relationship is established, by combining \textit{JWST}, \textit{Ariel} and \textit{HWO} phase curve datasets, we will have a sample of objects covering orbital periods to 10 -- 20 days and beyond, which will extend across the limit of tidal locking and answer our science objectives.

For each of these parameters there are also the following that will likely match the current state-of-the-art but need to be considered as important parameters to hold known to assess trends. 
\begin{itemize}
    \item Stellar spectral type: F,G,K,M, stars however recommendation is to focus on planets around F,G,K stars as large gas giants around M stars likely represent a different population. 
    \item Mass/gravity range of planets explored: state-of-the-art observations are already able to access the full range of masses for exoplanets. This study would focus on gas giant planets with large H/He fractions and therefore masses $>$0.4 Jupiter masses. In this mass range the average gravity is $\sim$15\,ms$^{-1}$, with current phase curves focused on the best short-period close-in targets which span a wide range from 6\,--\,40\,ms$^{-1}$. As we move to longer-period planets a careful sample of gravities will need to be assessed to better evaluate its effects on the measurable atmosphere. 
    \item Eccentricity: While a number of planets have been measured at different eccentricities, thus far they represent the extremes of the population e.g., circularized hot Jupiters, or highly eccentric giant planets on cometary like orbits. As planets move beyond the tidal locking radius we can expect that they will also have non-circular orbits. Therefore the eccentricity of the planet can be used to assess the likelihood of them being tidally locked to the star for a given orbital period. While there is no defined incremental, substantial or major progress to be made on the measurement of eccentricity it will need to be carefully considered for the population study and interpretation of the results. 
\end{itemize}

\section{Description of Observations}

\begin{table*}[ht!]
    \centering
     \caption[Observation Requirements]{Observational requirements for transiting exoplanets with \textit{HWO}.}
    \label{tab:obsreq}
    \begin{tabular}{|p{0.15\linewidth}|p{0.2\linewidth}|p{0.15\linewidth}|p{0.15\linewidth}|p{0.2\linewidth}|}
        \noalign{\smallskip}
        \hline
        \noalign{\smallskip}
        {Observation} & {State of the Art} & {Incremental Progress} & {Substantial Progress} & {Major Progress} \\
        \noalign{\smallskip}
        \hline
        \noalign{\smallskip}
        Amount of time on a single target &
        \textit{JWST}: Up to 60 hours continuous observation. \textit{Ariel}: Potentially up to a 5 day (120 hr) phase curve for select planets. &
        Up to 250 hour ($\sim$10 days) per target open up a wide range of close-in planet studies &
        Up to 500 hours ($\sim$20 days) per target &
        Over 500 hours per observation to observe the continuous phase curve of non-tidally locked planets on 20+ day orbits. \\
        \hline
        Wavelength Range &
        0.2 -- 0.8\,$\upmu$m with \textit{Hubble} WFC3/UVIS or STIS low resolution (R$\sim$20--70) STIS eschelle (R$\sim$115,000)
        \textit{JWST} from 0.6 -- 12\,$\upmu$m not simultaneous
        (R$\sim$20--2600) &
         0.15 -- 1.6\,$\upmu$m simultaneous (not necessarily continuous) low resolution R$\sim$200--2000 
        And R$>$150,000 eschelle over broad range &
         0.1 -- 2.5\,$\upmu$m  simultaneous medium resolution (R$\sim$3000--5000)
        with polarimetry (0.5 -- 1.0\,$\upmu$m) &  
        Continuous UV to IR coverage 0.1-- $>$5\,$\upmu$m at R$>$6,000 with polarimetry (0.3 -- 1.0\,$\upmu$m) \\
        \hline
        Spectroscopy and polarimetry &
        Low/medium resolution spectra R$\sim$20--2600 from space, R$>$100\,K on ground &
        Medium resolution spectra R$>$2000 at UV wavelengths &
        Medium resolution R$>$3000 across all wavelengths &
        Optical Polarimetry \\
        \hline
        Detectors &
        Observations are currently achieved by combining information from lots of detectors &
        Simultaneous observation with as few detectors as possible to cover the full $\lambda$
        Or, a single detector from UV to IR &
        Single photon counting UV, with separate standard IR detector &
        Single photon counting detectors \\
        \hline
        Magnitude of target in chosen bandpass &
        \textit{JWST}, V mag range 9th -- 15th magnitude obtained so far &
        V mag 5 -- 15th &
        Vmag 5 -- 18th &
        V mag range from 2 to 20th magnitude \\
        \hline
        SNR and Cadence of observations &
        \textit{JWST}, $\sim$30 ppm &
        5--10 ppm &
        100 ppb & 
        1 ppb \\
        \hline
        Pointing stability & 
        \textit{JWST} 0.1 pixel & 
        Within a tenth of a pixel & 
        Within 1000th of a pixel  &
        Absolute pointing combined with stable detectors \\
        \hline
        Pixel stability/detector thermal stability &
        Thermal stability. \textit{Hubble} has thermal fluctuation around earth orbit. \textit{JWST} is partially affected by electronics &
        Trackable fluctuations on the sub-pixel and sub-exposure level &
        Active subtraction of fluctuations using thermally adaptive systems &
        No fluctuations in the detectors (unrealistic)\\
        \hline
        Field of regard &
        \textit{JWST} - limited by solar elongation constraint to between 85 and 135 degrees due to sunshield.  HST limit is 55 degrees and is affected by gyroscopic pointing limitations. &
         &
         &
        All sky accessibility throughout the year\\
        \noalign{\smallskip}
        \hline
    \end{tabular}
\end{table*}

\textit{Hubble}, \textit{JWST}, along with dedicated missions such as \textit{Ariel}, will have made significant progress in characterizing transiting exoplanets, in particular those closest to their stars. However, phase curve observations will remain limited due to the timing required on a single target, the precision and resolution needed to assess their atmospheric profile, and the lack of spectroscopic coverage in the UV-optical to capture the reflected light from these worlds. The next big step in transiting exoplanet understanding will require long-term monitoring of individual systems to measure their full orbital period, to high precision, over multiple wavelengths, in order to empirically ascertain the tidal locking radius and impact of rotational dynamics on the atmosphere. Critically in the coming decades we will not be able to reach these goals in the UV-Optical at high precision and with the needed spectroscopic coverage. \textit{HWO} is the only telescope that can measure the reflected light of these planets, revealing their cloud structure, metal lines, and deep thermal profiles. 

Phase curves offer the best way to spatially sample transiting planets through measuring the full orbit covering two eclipse events and the primary transit. For continuous time series observations we require stable pointing of the telescope with limited physical interruption during the course of up to $\sim$50\,--\,500+ hours. These observations also require particular start times which will need to be accounted for in the schedule. Broad spectroscopic wavelength coverage is key to capture the transition from reflected to thermal components of the spectrum - if all wavelengths can be captured simultaneously you can mitigate the impact of changes to the telescope or planetary environment between observations and improve observatory efficiencies. Often for transiting planets they orbit around bright stars so a brightness limit to access V magnitude $\sim$5\,--\,20 (this encompasses HD\,189733 down to TRAPPIST-1 like stars and their planetary systems).

Spectroscopic observations are critical to measure atomic and molecular lines in the UV-optical and disentangle contributions from aerosol species and escaping atmospheres (see SCDD dos Santos). Low to medium resolution spectroscopy with simultaneous wavelength coverage adds incremental progress to current facilities such as \textit{Hubble} and \textit{Ariel}. Pushing to higher resolutions exceeding R\,$\sim$\,3000 in the UV-optical would enable the use of cross-correlation techniques \citep[e.g.,][]{BrogiLine2019AJ,Esparza2023ApJ} to identify atmospheric constituents. However, the extent of a single resolution element on the detector will be an important factor in setting the attainable resolution for the planetary spectrum \citep[e.g.,][using the R\,=\,2600 instrument to produce an R\,=\,100 planetary spectrum]{Espinoza2023PASP}. Higher resolutions still are required to explore spectral line broadening which can be related to rotational dynamics or atmospheric winds. Ground-based facilities in the IR exceeding R\,$\sim$\,25,000 have the potential to measure the rotational broadening of a directly imaged planet with a rotation period of $\sim$8 hours, however, resolutions exceeding R\,$\sim$\,100,000 are necessary for longer (thus slower) rotation periods (see SCDD Cubillos).

\section{Work that needs to be done in advance of HWO}
Transiting exoplanet characterization is an incredibly active field of investigation both theoretically and observationally. In the coming decades we can expect our knowledge of these worlds to have improved or changed substantially since the instigation of this document. Alongside these efforts a number of areas of study are required to provide predictions for \textit{HWO}. This is a non-exhaustive list of work that is required and we recommend that the community be called on to better define the steps that will need to be taken in the coming decades to establish the necessary information to answer our posed science goal. 

\subsection{Laboratory studies}
\begin{itemize}
    \item Aerosol properties in the UV-IR
    \item Lab measurements of optical properties of clouds inclusive polarimetry (e.g., Hamill, et al., 2024)
    \item Particle size distributions and aerosol fundamental structure
    \item Chemical networks for the production of different gases
    \item Sulfur and Phosphorus bearing species and their role in photochemistry
    \item Chemical relaxation timescales
\end{itemize}

\subsection{Modeling efforts}
\begin{itemize}
    \item GCM models of varied complexity for long period transiting exoplanets to provide a foundation for theoretical predictions in circulation.
    \item Signal-to-noise calculations in the UV-optical to determine at what resolution detections of line broadening from rotation or winds can be measured. 
\end{itemize}

\subsection{Phase Curve}
\begin{itemize}
    \item Ariel has the potential to substantially add science knowledge to the population of optical photometric phase curve studies. 
    \item It would be good to use some selected hot Jupiters as test cases for reflected light? Because a) there are still a lot of unknowns that \textit{HWO} can resolve and b) we can ideally benchmark the pipeline. Because here, we know the radius and thus can rule out some degeneracies. 
\end{itemize}

\subsection{Polarimetry}
\begin{itemize}
    \item Improvements of instruments to reach the few ppm levels.
    \item Better theoretical models to assess the parameter space for exoplanet studies with polarimetry. 
    \item In-depth models of polarimetric phase curves across the target parameter space for various cloud scenarios. 
    \item At these levels of polarimetry it is important to think of polarimetry in the design of \textit{HWO}.
\end{itemize}

{\bf Acknowledgements.}  
This work would not have happened without the additional support from members of the Characterizing Exoplanet Steering Committee and the \textit{HWO} START team. We thank Courtney Dressing and Pin Chen for their review of the science case. We thank the HWO administrative team for their work on supporting these science cases and their dissemination to the scientific community.
H.R.W was funded by UK Research and Innovation (UKRI) under the UK government’s Horizon Europe funding guarantee as part of an ERC Starter Grant [grant number EP/Y006313/1].

\bibliography{author.bib}

\end{document}